\documentclass{raa_twocolumn}           

\usepackage{graphicx,times}
\usepackage{natbib}
\usepackage{amssymb,amsmath}
\bibpunct{(}{)}{;}{a}{}{,}
\usepackage{wrapfig}
\usepackage[pagebackref=true]{hyperref}

\begin{document}

   \title{Timing and Spectral Analysis of 4U 1626-67 Using {\it \textbf {AstroSat}}/LAXPC}

 \volnopage{ {\bf 20XX} Vol.\ {\bf X} No. {\bf XX}, 000--000}
   \setcounter{page}{1}

   \author{N.J Juris
   \inst{1}, Marykutty James\inst{1}, Jincy Devasia\inst{2}
   }

   \institute{St.Thomas College, Ranni, Pathanamthitta - 689673, India; {\it njjuris@gmail.com}\\
	\and
        Henry Baker College, Melukavu, Kottayam - 686652, India
        \\
\vs \no
   {\small Received 20XX Month Day; accepted 20XX Month Day}
}

\abstract{ We present a comprehensive analysis of {\it AstroSat}/LAXPC data of the second spin-up and second spin-down phases of the persistent X-ray pulsar 4U 1626-67. Flares followed by a broad dip are detected in the spin-up observations. The pulse profiles changed from a shoulder-like structure to a broad sinusoidal shape as the source underwent a torque reversal from spin-up to spin-down. Energy-resolved pulse profiles in lower energies showed a double-horned profile in the spin-up state and a flat top with multiple peaks in the spin-down state. Regardless of the torque state, the pulse profiles exhibit a broad single-peaked shape at higher energies.
The observation in the spin-down era is characterised by the presence of a prominent QPO at $46.5\pm1.0$ mHz frequency. The QPO rms and centre frequency show a correlation with energy.
Spin-up and spin-down states show a difference in the shape of the power density spectrum.
After the torque reversal, a gradual flux drop and the hardening of the spectra were observed. 
The difference in the shape of the pulse profiles and the presence and absence of QPOs can be explained by the change in accretion flow geometry of the pulsar, from pencil-beam to fan-beam, between spin-down and spin-up states. 
 }

   \authorrunning{N.J Juris et al. }            
   \titlerunning{Timing and Spectral Analysis of 4U 1626-67}  
   \maketitle

%
\section{Introduction}           
\label{sec:intro}

Ultracompact X-ray binaries (UCXB) are a subgroup of low-mass X-ray binaries (LMXBs) with less than one-hour orbital periods. Since UCXBs have short orbital periods, only an evolved compact donor could fit into such tight orbits \citep{Tutukov1993ARep, Deloye2005ApJ}. The evolved UCXB donors are most likely anticipated to be hydrogen-deficient \citep{Paczynski1981ApJ}. The persistent LMXB pulsar 4U 1626-67 is classified as a UCXB due to its 42 min orbital period. This accreting X-ray pulsar located at a distance of $\sim$ 5-13 kpc from the Sun \citep{Chakrabarty1998ApJ}, is with a very low mass companion, with the estimated mass in the $\sim$ 0.03-0.09 $M_{\odot}$ range \citep{Levine1988ApJ}.  
It was discovered with the Uhuru satellite \citep{Giacconi1972ApJ}, and X-ray pulsations with a period of $\sim$ 7.7 s were located with SAS-3 observations by \cite{Rappaport1977ApJ}. 

Since its discovery, the persistent X-ray source 4U 1626-67 has experienced three torque reversals \citep{Camero-Arranz2010ApJ, Sharma2023MNRAS}. It was initially observed in a spin-up state and began to spin down in 1990 following a torque reversal. In 2008, a transition occurred from a consistent spin-down phase that lasted almost 18 years to a spin-up phase \citep{Benli2020MNRAS}. Again, it underwent a recent torque reversal in 2023 after 15 years of spin-up phase \citep{Sharma2023MNRAS}. 
The torque reversal is accompanied by changes in the luminosity of 4U 1626-67. The transition from the first spin-up state to the first spin-down state witnessed a decrease in the X-ray flux \citep{Chakrabarty1997ApJ}. The second torque reversal to the spin-up phase was accompanied by an increase in the source luminosity about two to three times \citep{Jain2010MNRAS}. In 2010, two years following that torque reversal, the source's X-ray flux reached almost the same level as in 1977 \citep{Camero-Arranz2010ApJ}. The recent shift to a spin-down state in 2023 is accompanied by a decrease in the luminosity of the source, similar to the first detected phase transition \citep{Jenke2023ATel, Sharma2023MNRAS}.

Both X-ray and optical flares have been observed in this LMXB pulsar since the earliest observations \citep{McClintock1980ApJ}, with detections continuing consistently over the following years \citep{Raman2016MNRAS, Beri2018MNRAS}. 4U 1626-67 shows strong flaring activity with a duration of a few hundred seconds during the spin-up state and comparatively fewer flares in the spin-down state \citep{Chakrabarty2001, Beri2014MNRAS}. X-ray flares of different durations and recurrence times have been previously reported in many other sources like 4U 1901+03 \citep{James2011MNRAS}, SMC X-1, and LMC X-4 \citep{Moon2003ApJ_smcx-1, Moon2003ApJ_lmcx-4}.
The X-ray characteristics of 4U 1626-67 differ in spin-up and spin-down phases. A notable difference is the disappearance of the bi-horned peaks observed in the low-energy pulse profiles during the spin-up state, from the profiles of the spin-down state \citep{Beri2014MNRAS, Beri2018MNRAS}. 

Quasi-Periodic Oscillations in the power spectrum of accretion-powered X-ray pulsars, are believed to arise due to inhomogeneities in the inner accretion disc and are considered as an important signature of the presence of an accretion disc.
During the spin-down eras of 4U 1626-67, observations revealed prominent QPOs around 48 mHz, displaying a gradual evolution in frequency over time \citep{Chakrabarty1998ApJ}. 
In the first spin-down phase, QPOs at 48 mHz were consistently observed across all observations, while these QPOs were absent during the following spin-up phase \citep{Kaur2008ApJ, Jain2010MNRAS}. However, weak and broad QPOs around 40 mHz were identified in the initial spin-up phase from {\it Ginga} observations, indicating variations in QPO characteristics between spin-up and spin-down states \citep{Shinoda1990PASJ}. These aperiodic variabilities are detected in several other X-ray sources including KS 1947+300 \citep{James2010MNRAS}, LMC X-4 \citep{Moon2001ApJ}, and Cen X-3 \citep{Raichur2008ApJ}.

The spectrum of 4U 1626–67 has been extensively studied using data from various observatories, revealing a two-component structure composed of a blackbody and a power law \citep{Pravdo1979ApJ, Kii1986PASJ, Krauss2007ApJ, Camero-Arranz2012A&A}. These continuum spectral parameters have been observed to vary during periods of torque reversal. 
The spectrum had a blackbody temperature of around 0.6 keV and a power-law photon index of about 1.5 in the first spin-up state \citep{Pravdo1979ApJ, Kii1986PASJ}.
The blackbody temperature dropped to about 0.3 keV during the first spin-down state, and the energy spectrum became comparatively harder with a power-law index of 0.4–0.6 \citep{Angelini1995ApJ, Orlandini1998ApJ}.
Similar to the first spin-up state, the second spin-up phase showed a higher blackbody temperature of roughly 0.5–0.6 keV but the photon index did not go back to the previous range and showed values near 0.8–1.0 \citep{Krauss2007ApJ, Camero-Arranz2010ApJ, Jain2010MNRAS}. 
Additionally, numerous spectroscopic observations have reported unusually bright emission lines of neon (Ne) and oxygen (O) \citep{Krauss2007ApJ, Camero-Arranz2010ApJ}.
The surface magnetic field of 4U 1626-67 is estimated to be approximately $\sim$ 3$\times$10$^{12}$ Gauss, based on the detection of a cyclotron line feature around 37 keV in its spectra, observed across multiple observatories during both spin-up and spin-down states \citep{Pravdo1979ApJ, Orlandini1998ApJ, Camero-Arranz2010ApJ}. 

This work includes the timing and spectral analysis of the X-ray binary 4U 1626-67 utilizing data from Large Area X-ray Proportional Counter (LAXPC) onboard {\it AstroSat} for the first time.
We have analysed the LAXPC20 observations of this source from January 2016 to May 2023. The observations and data analysis are described in section \ref{sec:Obsv} and timing analysis is carried out in section \ref{subsec:timing}. The pulse profiles and QPO are studied in detail in sections \ref{subsec:pulspro} and \ref{subsec:pds}. Section \ref{subsec:spec} describes spectral analysis followed by discussions and conclusions in section \ref{sect:discussion}.

\section{Observations and Data analysis}
\label{sec:Obsv}

{\it AstroSat}, the first multi-wavelength astronomical satellite from India, was launched in 2015 \citep {Singh2014SPIE}. Its five payloads are the Large Area X-ray Proportional Counter (LAXPC), Cadmium-Zinc-Telluride Imager (CZTI), Soft X-ray Telescope (SXT), Ultraviolet Imaging Telescope (UVIT), and Scanning Sky Monitor (SSM). The LAXPC is composed of three detectors, the LAXPC10, LAXPC20, and LAXPC30. LAXPC operates in the 3-80 keV energy range with 6000 cm$^{2}$ in area and a 10 ms time resolution. It operates in two modes: Broad Band Counting and Event Mode. This work uses the event mode data as it records the energy, identity, and arrival time of each event \citep{Yadav2016SPIE}.

{\it AstroSat} made several observations of 4U 1626-67 between January 2016 and May 2023. The observations are downloaded from the Indian Space Science Data Centre (ISSDC) archive \footnote{https://www.issdc.gov.in/astro.html} and the observational details are given in Table 1. In this work, we limit our research to data from LAXPC20 as the LAXPC30 detector is not working and LAXPC10 exhibits gain variation and gas leakage. For LAXPC data analysis, we used the tools available in LAXPCSoft V3.4.4 developed at TIFR/IUCAA \footnote{https://www.tifr.res.in/astrosat\_laxpc/LaxpcSoft.html}. Spectral analysis was done using the {\sc XSPEC} version 12.11.1 available in HEASOFT \footnote{https://heasarc.gsfc.nasa.gov/docs/software/lheasoft/download.html} version 6.28.

\begin{table*}
\centering
    \caption{Details of {\it AstroSat}/LAXPC observations of 4U 1626-67 on different epochs.} \label{tab:obsv}
    \begin{tabular}{lcccc}
        \hline \\[1pt]
        Data & Obs. ID & Observation date  & No.of orbits & Exposure time (ks) \\[5pt]
        \hline  \\[1pt]
        Data1 & G02\_020T02\_9000000268 & 2016 January 13 & 4 & 8.21 \\[5pt]
        Data2 & G02\_020T02\_9000000294 & 2016 January 26 & 5 & 12.19 \\[5pt]
        Data3 & G05\_021T01\_9000000624 & 2016 August 26 & 12 & 41.69 \\[5pt]
        Data4 & G07\_049T01\_9000001352 & 2017 July 2 & 21 & 67.85 \\[5pt]
        Data5 & G08\_084T01\_9000002100 & 2018 May 15 & 23 & 36.35 \\[5pt]
        Data6 & T05\_106T01\_9000005642 & 2023 May 18  & 21 & 42.06 \\[5pt]
        \hline
    \end{tabular}
     \vspace{0.5cm}
\end{table*}

The LAXPC level-1 data are processed to extract events, light curves, spectra, and background files for the Good Time Intervals (GTI). The LAXPC pipeline software runs by taking the required input files from its well-defined directory structure, which includes level-1 data files, calibration, and background files for the generation of level-2 data products. The required background files are supplied along with the software. The software itself offers guidance for the selection of a suitable background as well as a response file that may be used for further analysis \citep{Mukerjee2021ApJ}.
The pipeline generates background-subtracted light curves and calibrated spectra for further analysis of the source. Light curves were corrected to the solar system barycentre using $barycorr$ \footnote{https://heasarc.gsfc.nasa.gov/docs/software/heasoft/help/barycorr.html} by HEASOFT.

\section{Results}
\label{sec:resul}
\subsection{Timing Analysis} \label{subsec:timing}
\subsubsection{Light Curves}\label{subsec:lcurve}

\begin{figure*}
   \includegraphics[width=0.45\textwidth, angle=0]{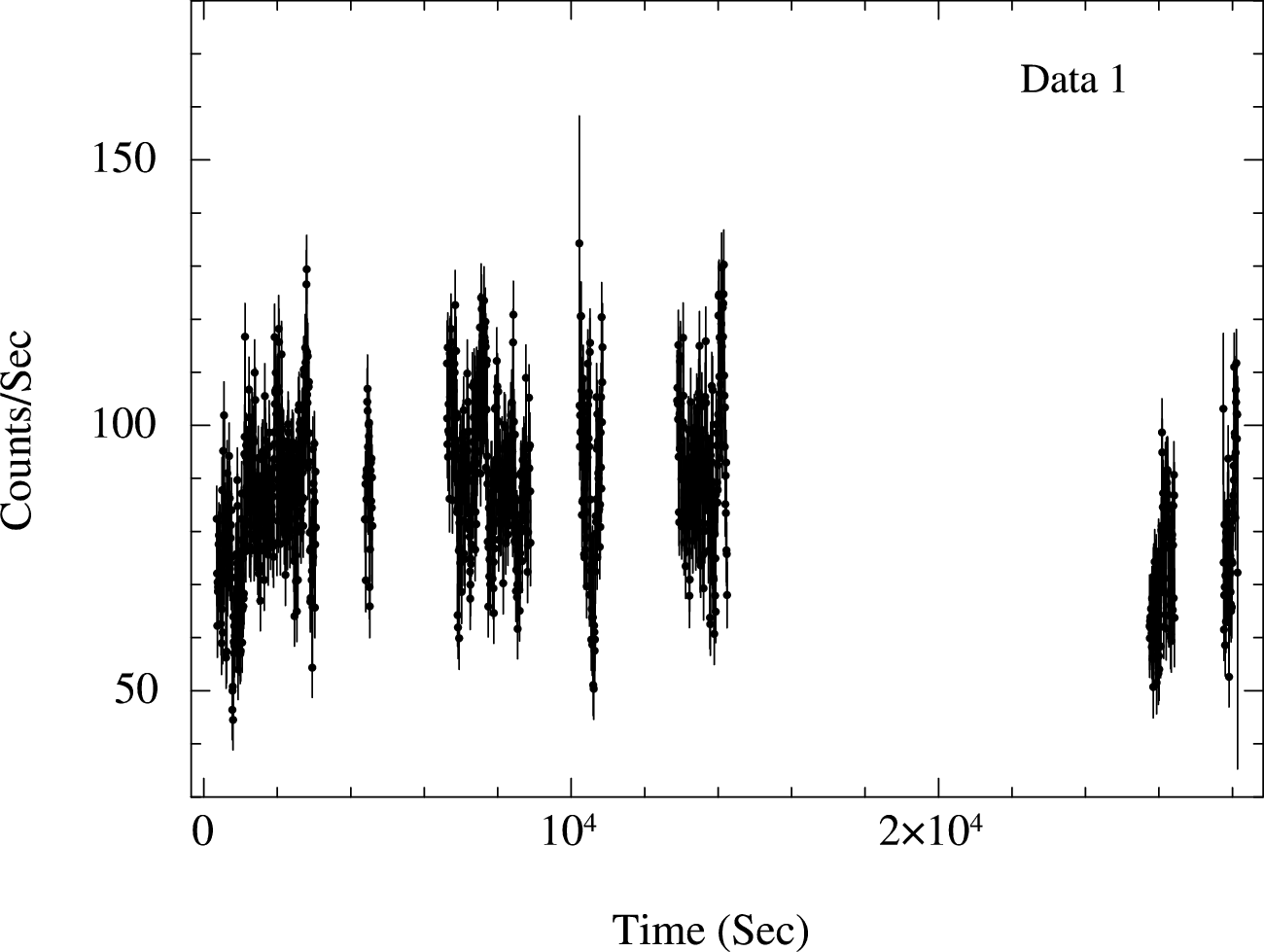}
   \hspace{1cm}
   \includegraphics[width=0.45\textwidth, angle=0]{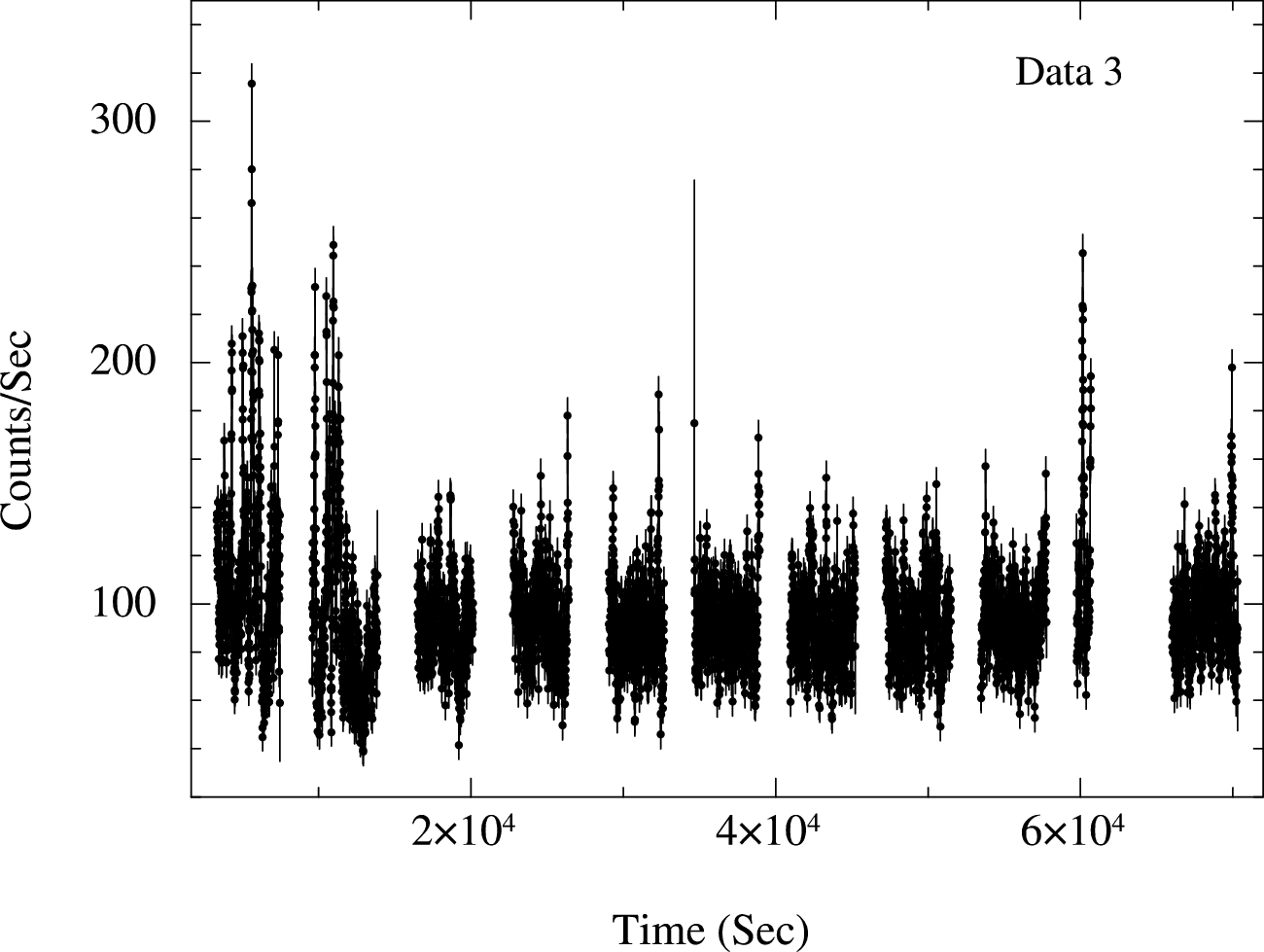}
   \caption{The 3-80 keV background-subtracted light curves of 4U 1626-67 from LAXPC observations of January 2016 (left) and August 2016 (right).} 
   \label{lc1}
\end{figure*}

\begin{figure*}
   \includegraphics[width=0.45\textwidth, angle=0]{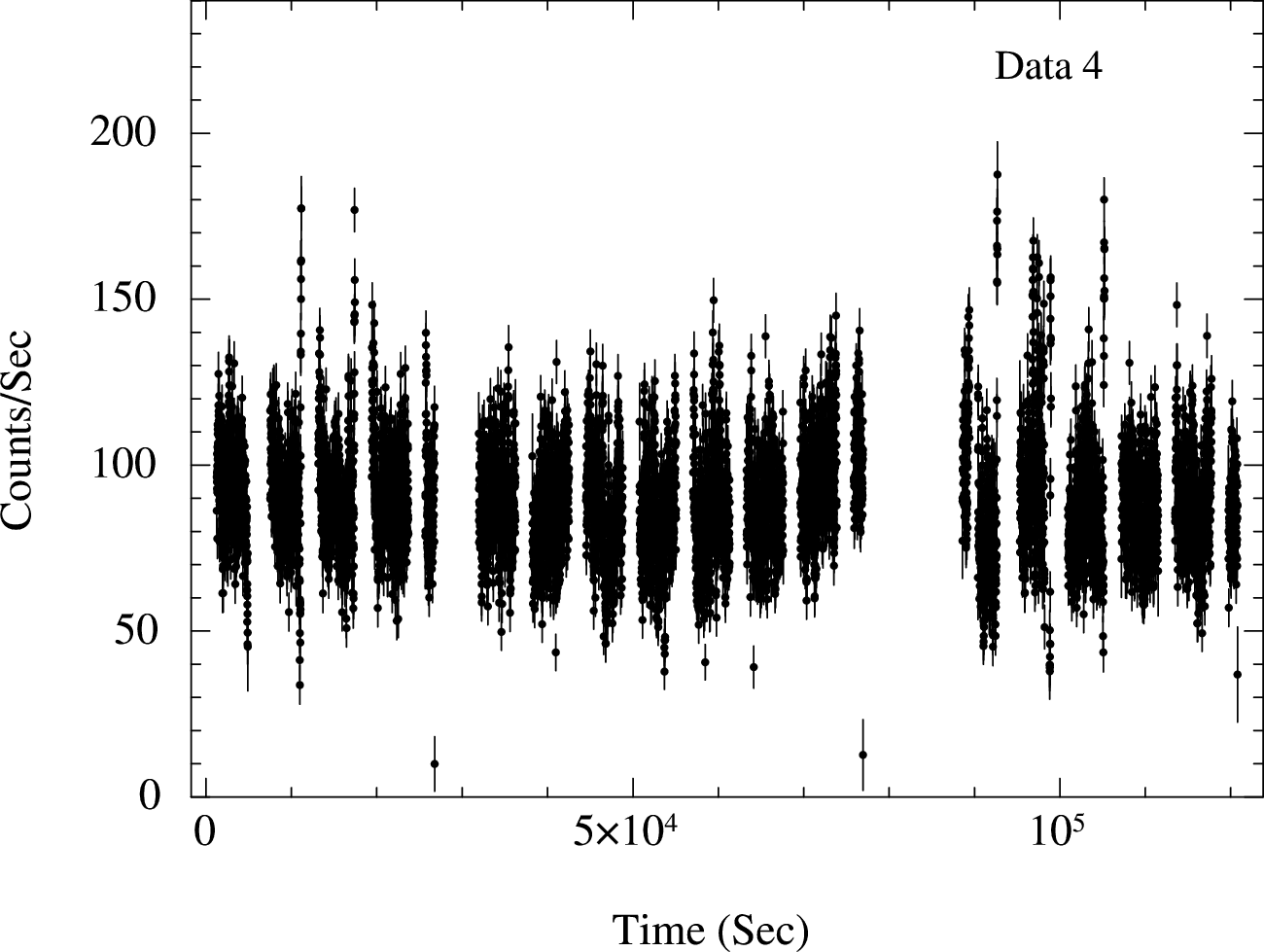}
   \hspace{1cm}
   \includegraphics[width=0.45\textwidth, angle=0]{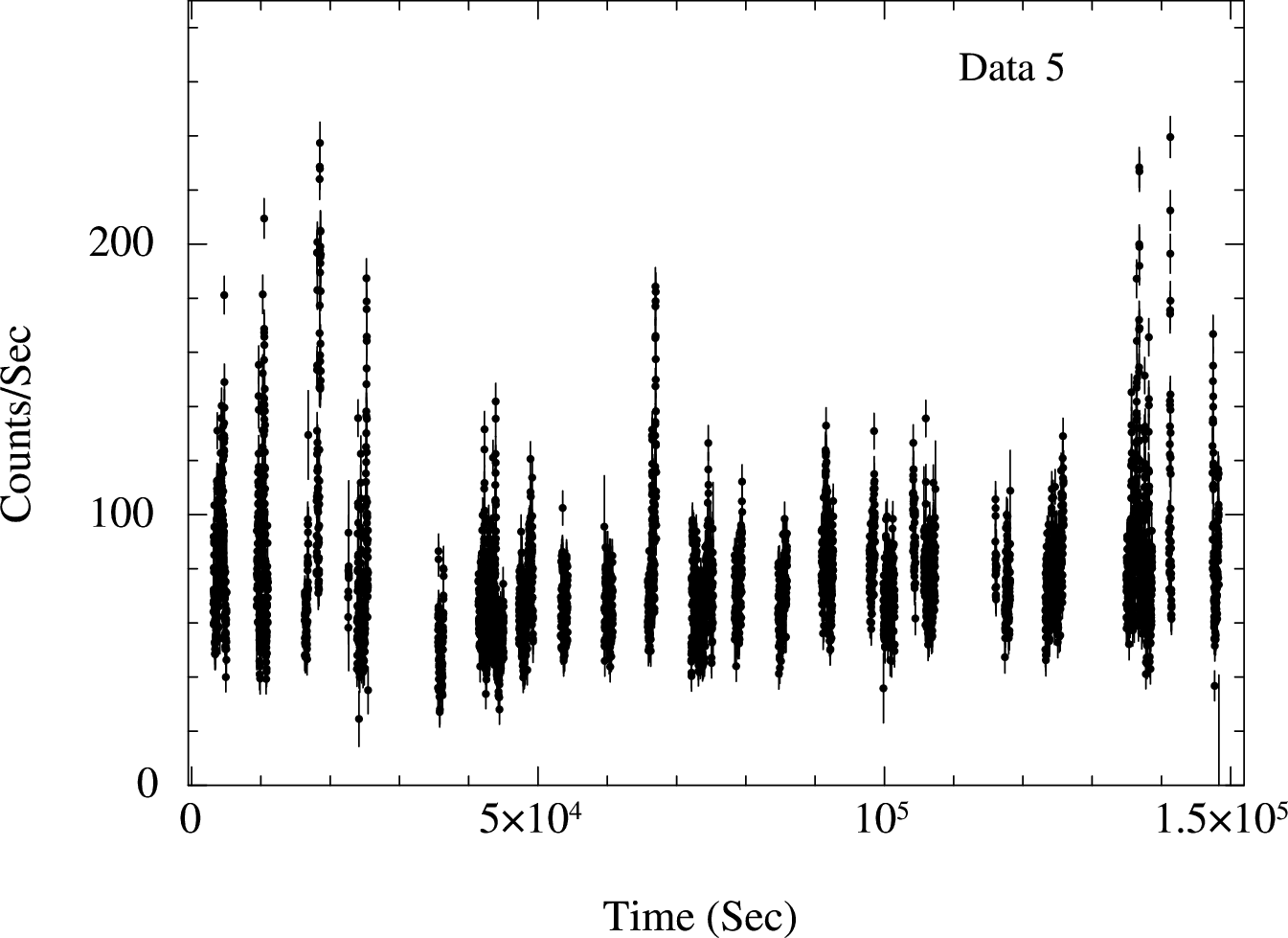}
   \caption{The 3-80 keV background-subtracted light curves of 4U 1626-67 from LAXPC observations of July 2017 (left) and May 2018 (right).} 
   \label{lc2}
\end{figure*}

\begin{figure*}
      \includegraphics[width=0.45\textwidth, angle=0]{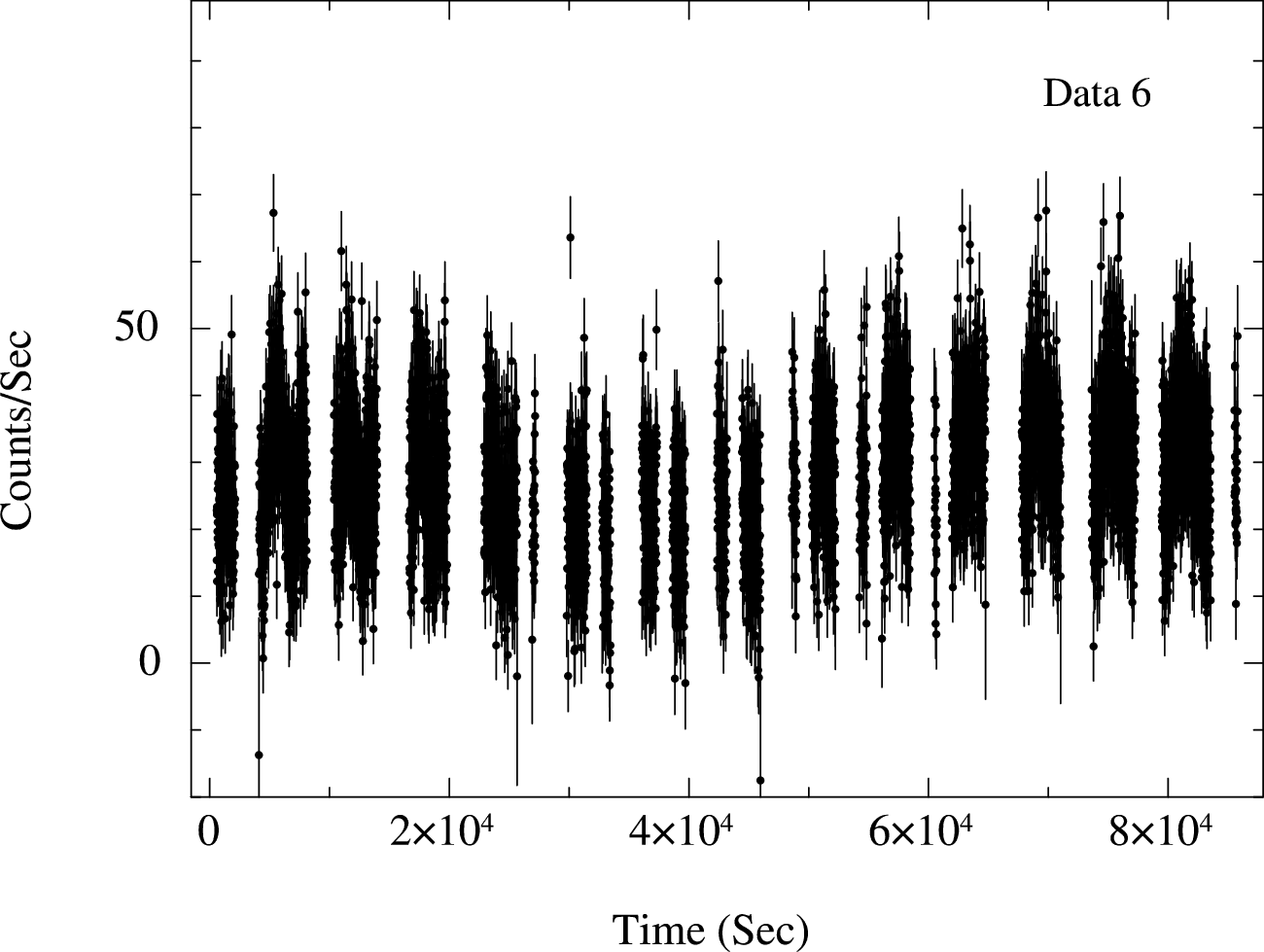}
   \hspace{1cm}
      \includegraphics[width=0.45\textwidth, angle=0]{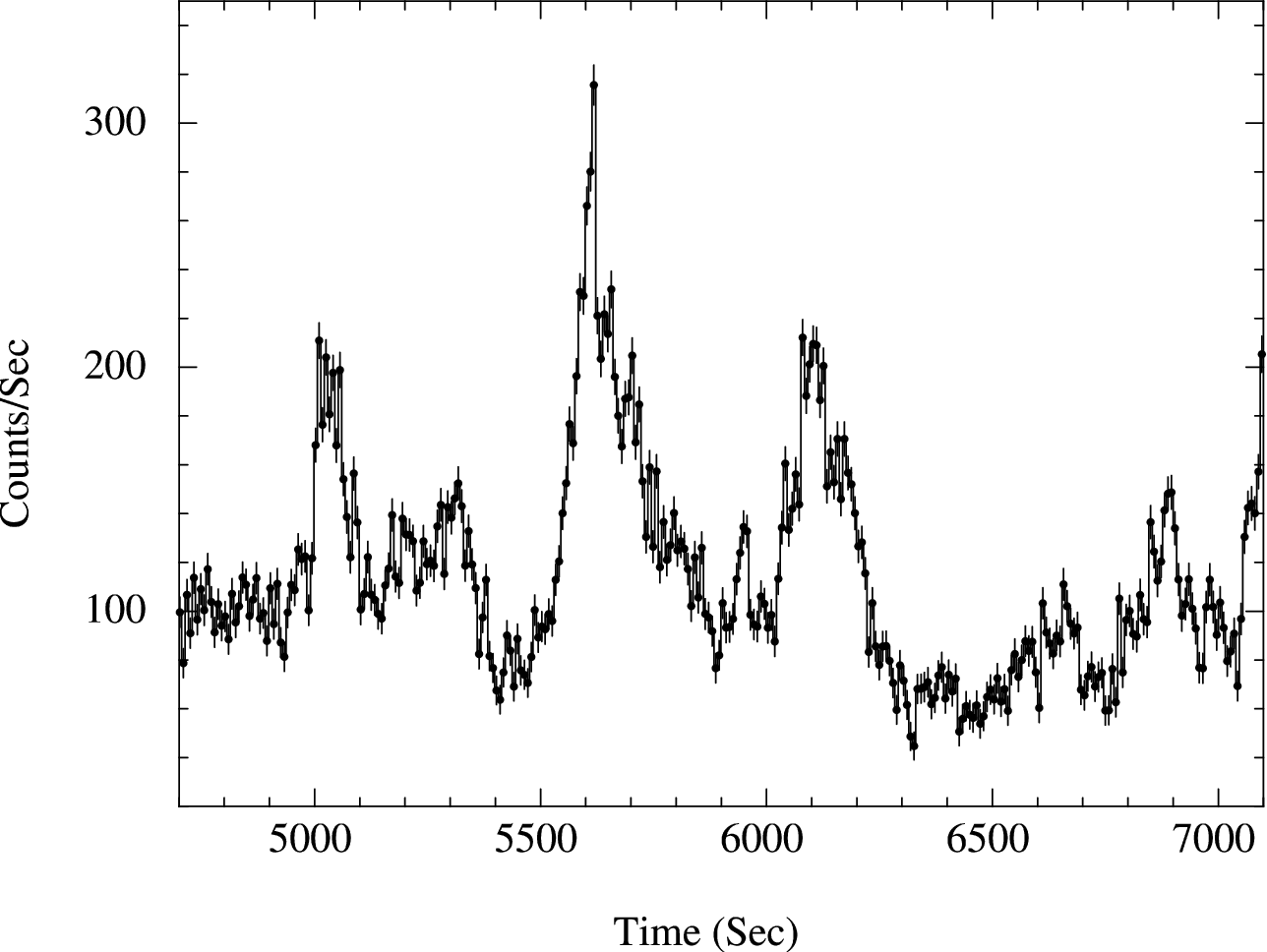}
    \caption{The 3-80 keV background-subtracted light curve of 4U 1626-67 from LAXPC observations of May 2023 (left) and the rescaled light curve of Data 3, showing the region with consecutive flares and the subsequent dip (right).}
   \label{lc3}
\end{figure*}

\begin{figure*}
      \includegraphics[width=0.45\textwidth, angle=0]{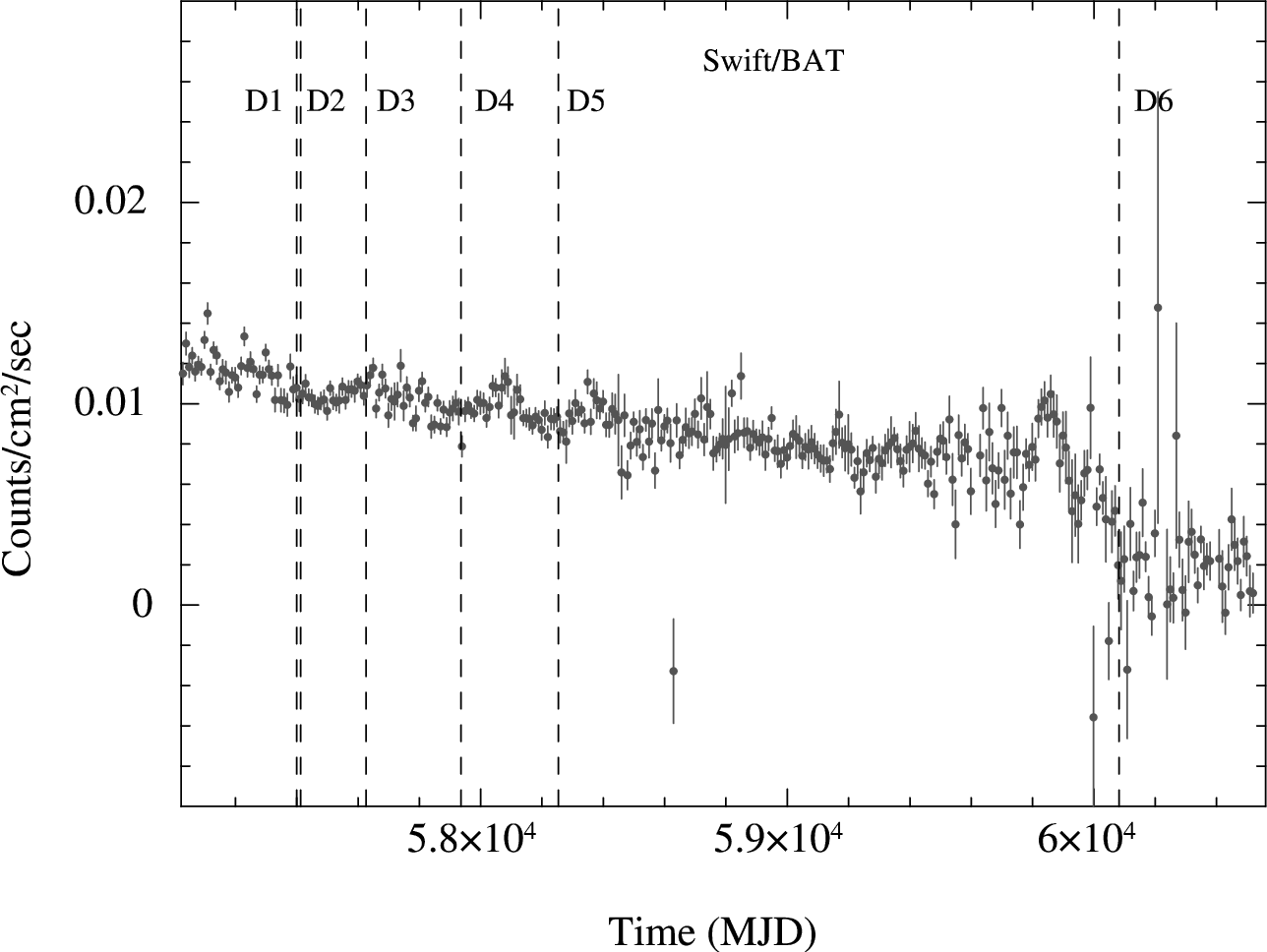}
    \hspace{1cm}
    \includegraphics[width=0.45\textwidth, angle=0]{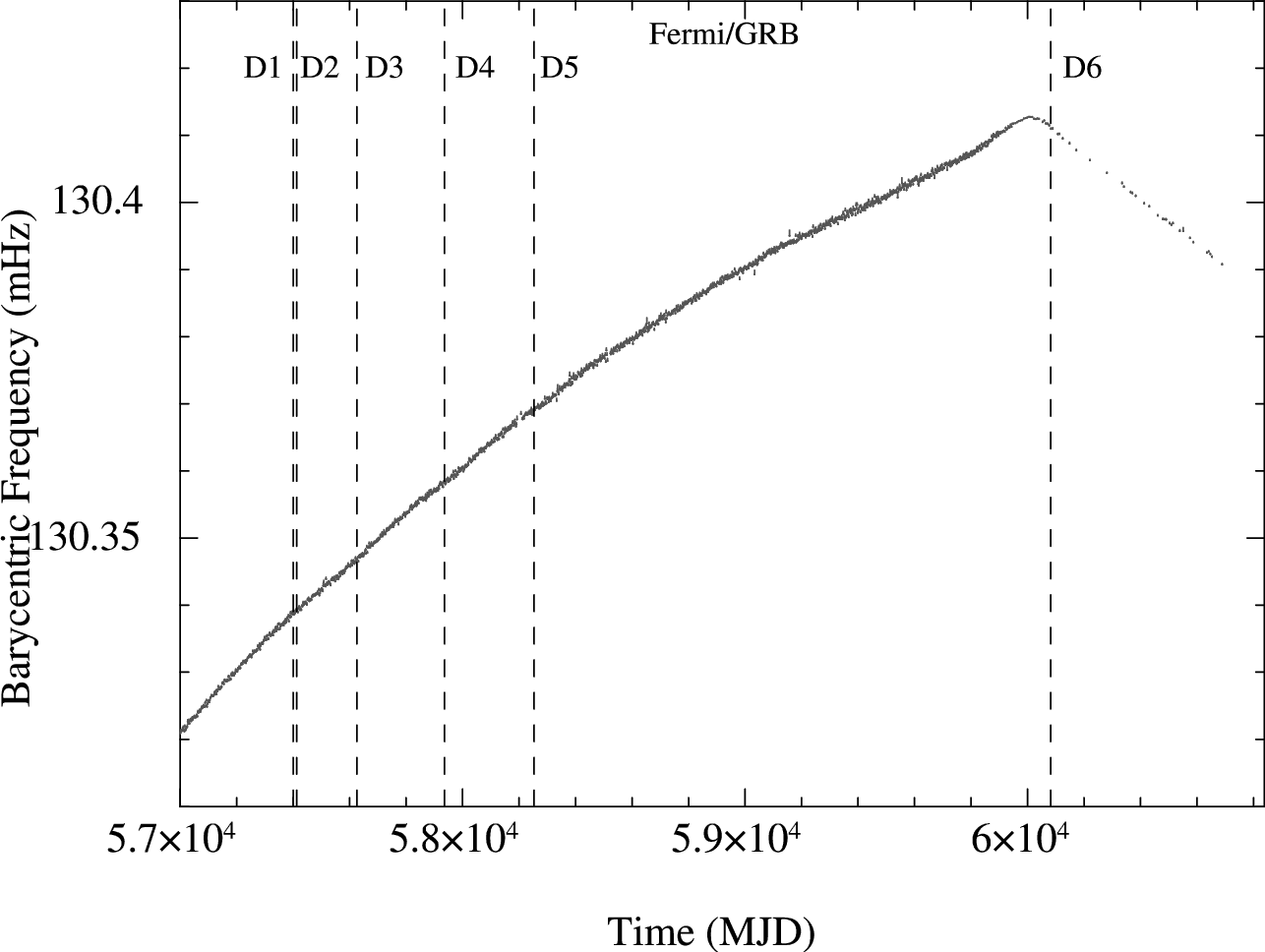}
    \caption{The {\it Swift}-BAT light curve of 4U 1626-67 in the 15-50 keV range, binned to 10 days (left) and the {\it Fermi}-GBM spin history of 4U 1626-67 (right). Vertical dotted lines on both plots denote the LAXPC20 observations.}
   \label{lc4}
\end{figure*}

We created 3-80 keV background-subtracted light curves using the main anodes and all layers of LAXPC20 for a bin size of 0.01 s. The source has a persistent count rate of $\sim$ 100 c s$^{-1}$ in Data 1 to Data 5 when it was in the spin-up state. The average intensity of the source seems to have been reduced to $\sim$ 30 c s$^{-1}$ in Data 6 when the source returned to spin-down.
The light curves of Data 3, Data 4, and Data 5 show the presence of flares. These flares have an intensity two or more times greater than the quiescent level and are 100 s to 300 s long. They have a symmetric shape with steady rise and decay and reoccur with time scales varying between 300 s and 700 s.  
The flares of 4U 1626-67 in the spin-up state of our observations, show energy dependence and are visible up to $\sim$ 15 keV in the light curves.
In the Data 3 light curve, two flares are seen together with $\sim$ 100 s separation; the first one being the largest flare in all the observations analysed in this work. The first and second flares have their intensity rising above $\sim$ 3 and $\sim$ 2 times the quiescent level, respectively. A broad dip reducing in intensity to below the persistent level was observed after the decay of the second flare at 6200 s of Figure \ref{lc1} (right panel). This dip in the intensity recovered to the persistent level after $\sim$ 250 s and is shown in Figure \ref{lc3} (right panel). The barycenter-corrected light curves of all the observations with a bin size of 7.7 s are shown in Figures \ref{lc1}, \ref{lc2}, \ref{lc3} and \ref{lc4}. 
The left panel of Figure \ref{lc4} shows the positions of LAXPC20 observations in the {\it Swift}- BAT light curve of 4U 1626-67, and the right panel shows the {\it Fermi}-GBM spin history of 4U 1626-67.

\subsubsection{Pulse Profile}\label{subsec:pulspro}

\begin{figure}
   \includegraphics[width=0.45\textwidth, angle=0]{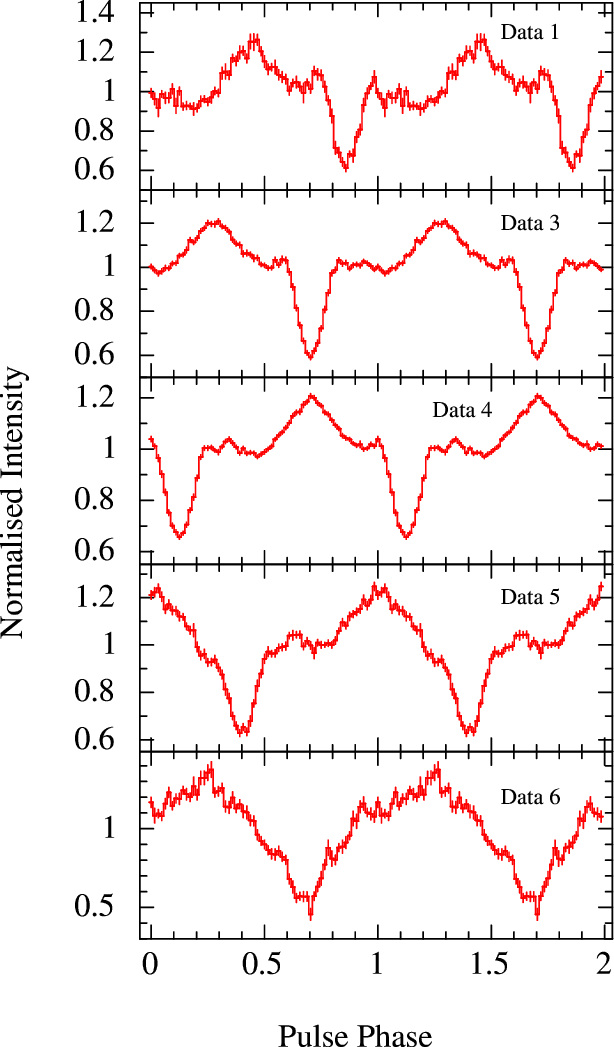}
   \caption{The 3-80 keV pulse profiles of 4U 1626-67 generated from observations made with {\it AstroSat}/LAXPC. The folded light curves are normalised to counts/s.} 
   \label{pulse_all}
\end{figure}

Using the FTOOLS task $EFSEARCH$, we applied the $\chi^{2}$ maximization method and pulse folding to obtain the spin period. Spin periods of 7.6723 s, 7.6723 s, 7.6718 s, 7.6712 s, 7.6705 s, and 7.6681 s were obtained respectively for Data 1, Data 2, Data 3, Data 4, Data 5, and Data 6.
The 3-80 keV light curves were folded at the respective spin periods. The pulse profile of Data 2 is distorted and with very few statistics, thus is omitted from analysis. Energy average pulse profiles with 64 phase bins created for all other observations are plotted in Figure \ref{pulse_all}.
3-80 keV pulse profile of Data 1 to 5 shows a main peak and two steps on both sides (a shoulder-like structure). The small step before the main peak is of lesser amplitude than the following step.
In Data 1, a significant dip in the intensity is observed before the main peak; which diminishes in the subsequent observations. 
The small peak after the main peak, broadens and eventually joins with the latter, as the source evolves in the spin-up state. 
The shift from spin-up to spin-down, between data 5 and data 6, of the source is characterised by the change in the shape of the energy-averaged pulse profile to a broad sinusoid with many spikes.

\begin{figure*}
   \centering
   \includegraphics[width=0.45\textwidth, angle=0]{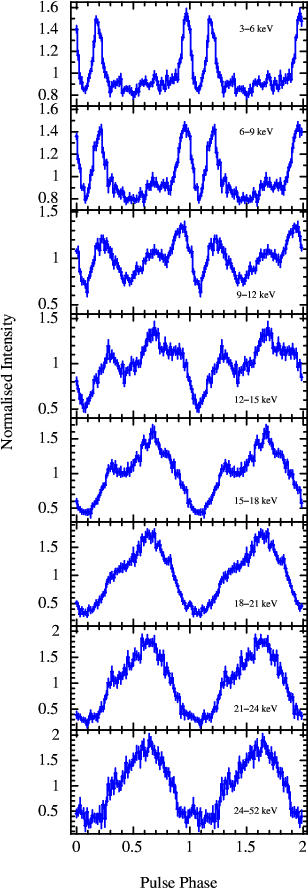}
   \hspace{1cm}
   \includegraphics[width=0.45\textwidth, angle=0]{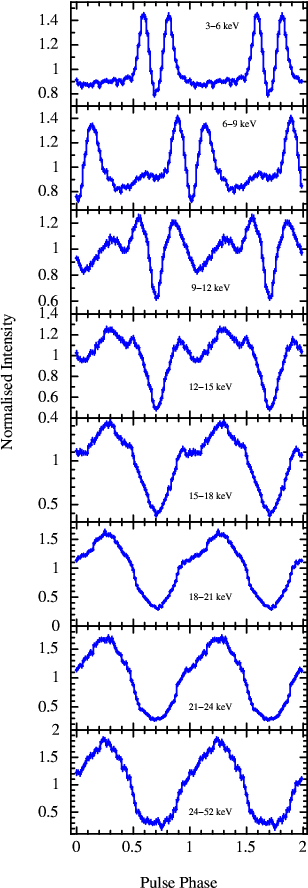}  
   \caption{Energy resolved pulse profile of Data 1 (left), and Data 3 (right) of 4U 1626-67 generated from observations made with {\it AstroSat}/LAXPC.} 
   \label{pulse1}
   \vspace{1cm}
\end{figure*}

\begin{figure*}
\centering
    \includegraphics[width=0.45\textwidth, angle=0]{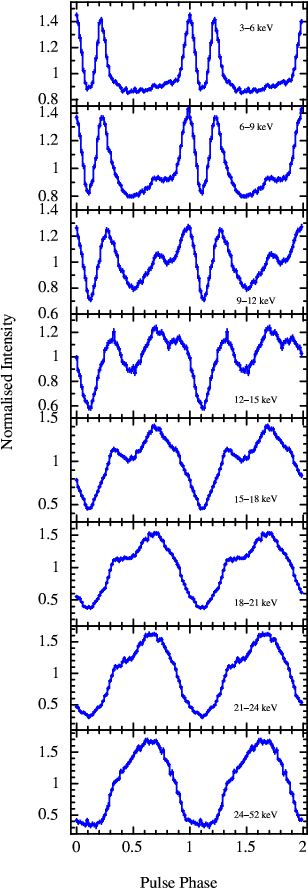}
   \hspace{1cm}
   \includegraphics[width=0.45\textwidth, angle=0]{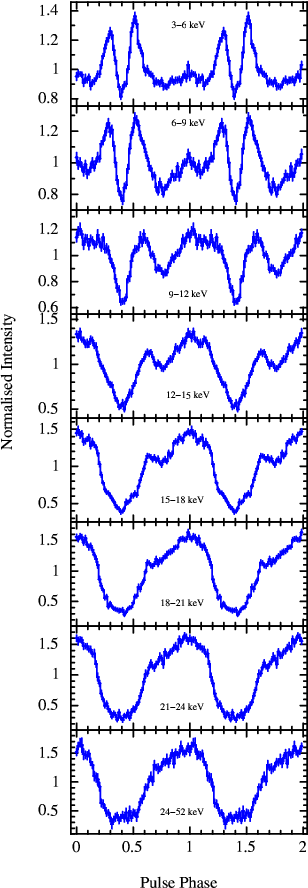}
   \caption{Energy resolved pulse profile of Data 4 (left), and Data 5 (right) of 4U 1626-67 generated from observations made with {\it AstroSat}/LAXPC.} 
   \label{pulse2}
\end{figure*}

\begin{figure}
\centering
   \includegraphics[width=0.45\textwidth, angle=0]{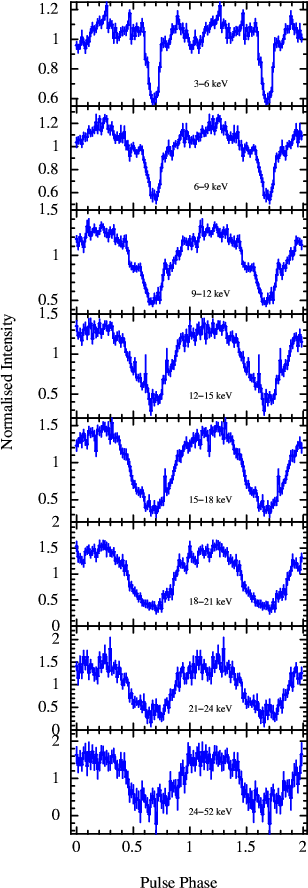}
 \caption{Energy resolved pulse profile of Data 6 of 4U 1626-67 generated from observations made with {\it AstroSat}/LAXPC.} 
   \label{pulse3}
\end{figure}

4U 1626-67 is known to show strong energy-dependent pulse profiles. We created pulse profiles using the light curves in the energy bands of 3-6 keV, 6-9 keV, 9-12 keV, 12-15 keV, 15-18 keV, 18-21 keV, 21-24 keV and 24-52 keV. We could detect pulsations only up to 52 keV, above that, pulsations were undetectable due to poor statistics in the data. These energy-dependent pulse profiles of all the observations are plotted in Figures \ref{pulse1}, \ref{pulse2} and \ref{pulse3}. 

The pulse profile during the spin-up state (Figures \ref{pulse1} and \ref{pulse2}) shows a double-horned pulse shape separated by a narrow dip, in the low energies and a single-peak profile in higher energies. In the lower energy bands up to 9 keV, a sharp double-peak is observed; the first peak being greater in amplitude than the second peak except in the case of Data 5 (Figure \ref{pulse2} right panel) where the second peak is of greater amplitude. With increasing energy, the narrow dip between the two peaks broadens. The phase before and after the double-peaks is characterised by the presence of a broad minimum; which disappears as the profile evolves in energy. Above 9 keV, a small peak appears just before the first peak and increases in intensity. No double-horned structure is observed above 18 keV, resulting in the shape change of the profile to a single-peaked pulse.

After the torque reversal, the pulse profiles changed in morphology to a flat top with many small peaks in the lower energies and a broad sinusoid in the higher energies as shown in Figure \ref{pulse3}. The 3-6 energy band shows a flat broad profile with multiple small peaks. The shape slightly shifts to a broad single peak in the following energies. Above 18 keV, the pulse profile has a broad sinusoidal shape with many spikes on it.
The pulse fraction (PF), which is expressed by the relation $PF=\frac {P_{max}-P_{min}} {P_{max}+P_{min}}$, represents the relative amplitude of the changing pulse profile. The PFs for Data 1, Data 3, Data 4, Data 5, and Data 6 are $34.8\pm1.6\%$, $34.4\pm0.9\%$, $29.6\pm0.6\%$, $33.3\pm1.1\%$ and $50.5\pm3.3\%$, respectively. 
It may be noted that there is a statistically significant increase in the value of PF as the source underwent torque reversal. While a single observation is not sufficient to definitively conclude this, the observed trend suggests an increase in PF as the source underwent a torque reversal from spin-up to spin-down. This trend aligns with previous studies of the source, which have reported similar behaviour during torque reversals \citep{Beri2014MNRAS, Sharma2023MNRAS}.

\subsubsection{Power Density Spectra}\label{subsec:pds}

Power Density Spectrum (PDS) was created using the FTOOL $POWSPEC$ for each data set to search for any aperiodic variability. The PDS was made for small data segments of 8192 s, and approximately 6 intervals were averaged into a single frame to improve the detectability of the QPO-like features. The PDS was normalized (norm= -2) such that its integral gives the squared RMS fractional variability, and the white noise level is subtracted.

\begin{figure}
    \includegraphics[width=0.45\textwidth, angle=0]{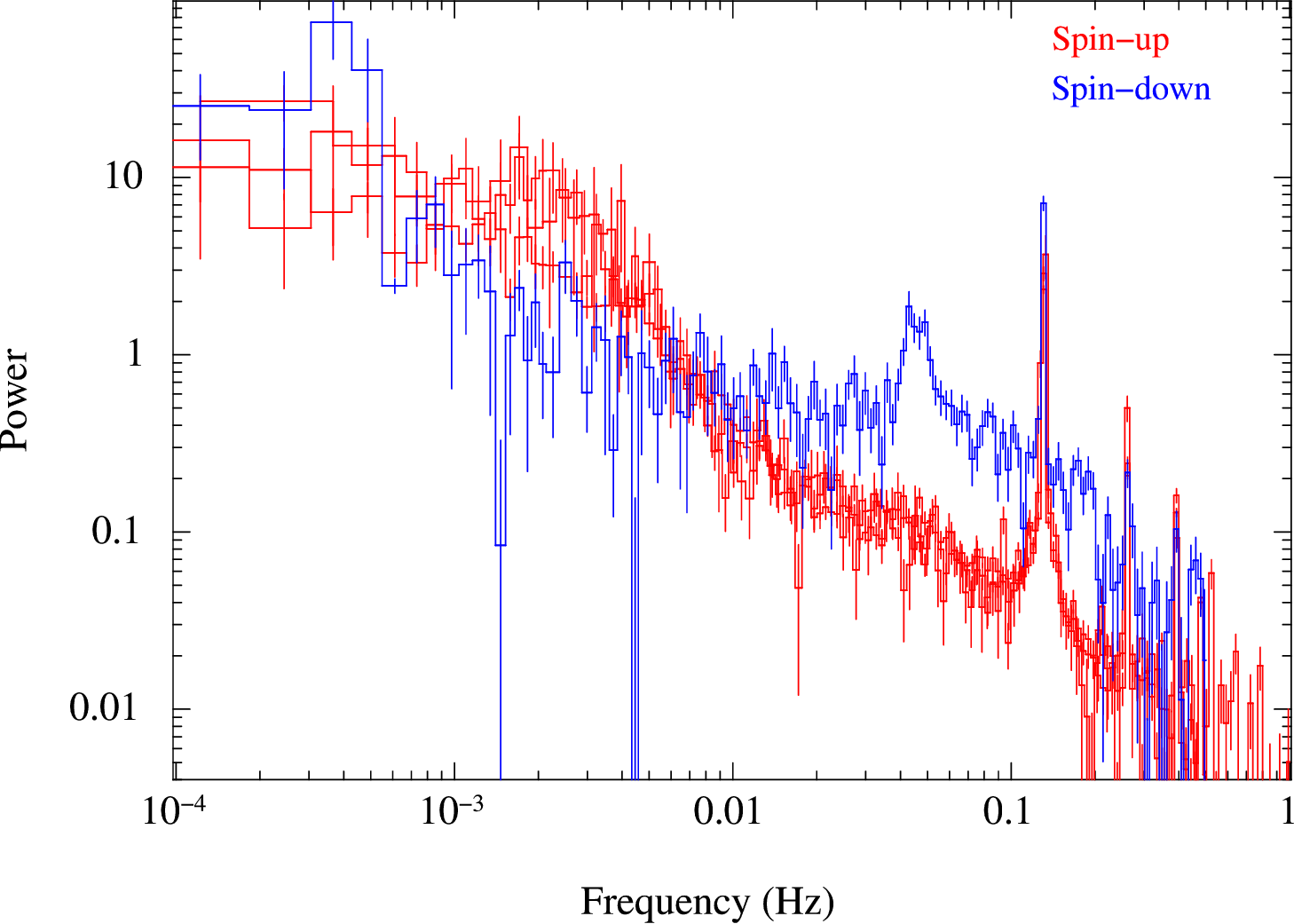}
    \caption{The 3-80 keV PDS of 4U 1626-67 from {\it AstroSat}/LAXPC observations of both the spin-up and the spin-down states.} 
    \label{pds_two}
\end{figure}

We have noticed a change in the shape of the PDS of our observations between the spin-up and spin-down states. Data 1 to Data 5 exhibited an inclined-shaped continuum with a broad bump around the $\sim$ 2 mHz frequency. This broad feature was energy-dependent and distinguishable up to $\sim$ 21 keV. Additionally, a broadening of the spin frequency peak was evident in the PDS of Data 3, Data 4, and Data 5. 
However, as the source transitioned into the spin-down phase following a torque reversal, the PDS continuum flattened, and both the broad feature at low frequencies and the broadening of the spin frequency peak disappeared. Notably, during the spin-down state (Data 6), a prominent QPO-like feature at $46.5\pm1.0$ mHz was detected. The same has been identified with $NuSTAR$ observation of the source during 2023 May by \cite{Sharma2023MNRAS}. The QPO feature in the PDS could be detected up to $\sim$ 15 keV.
The PDS of spin-up states (in red) and spin-down states (in blue) are overlayed and plotted in Figure \ref{pds_two} for comparison. 

\begin{figure}
    \includegraphics[width=0.45\textwidth, angle=0]{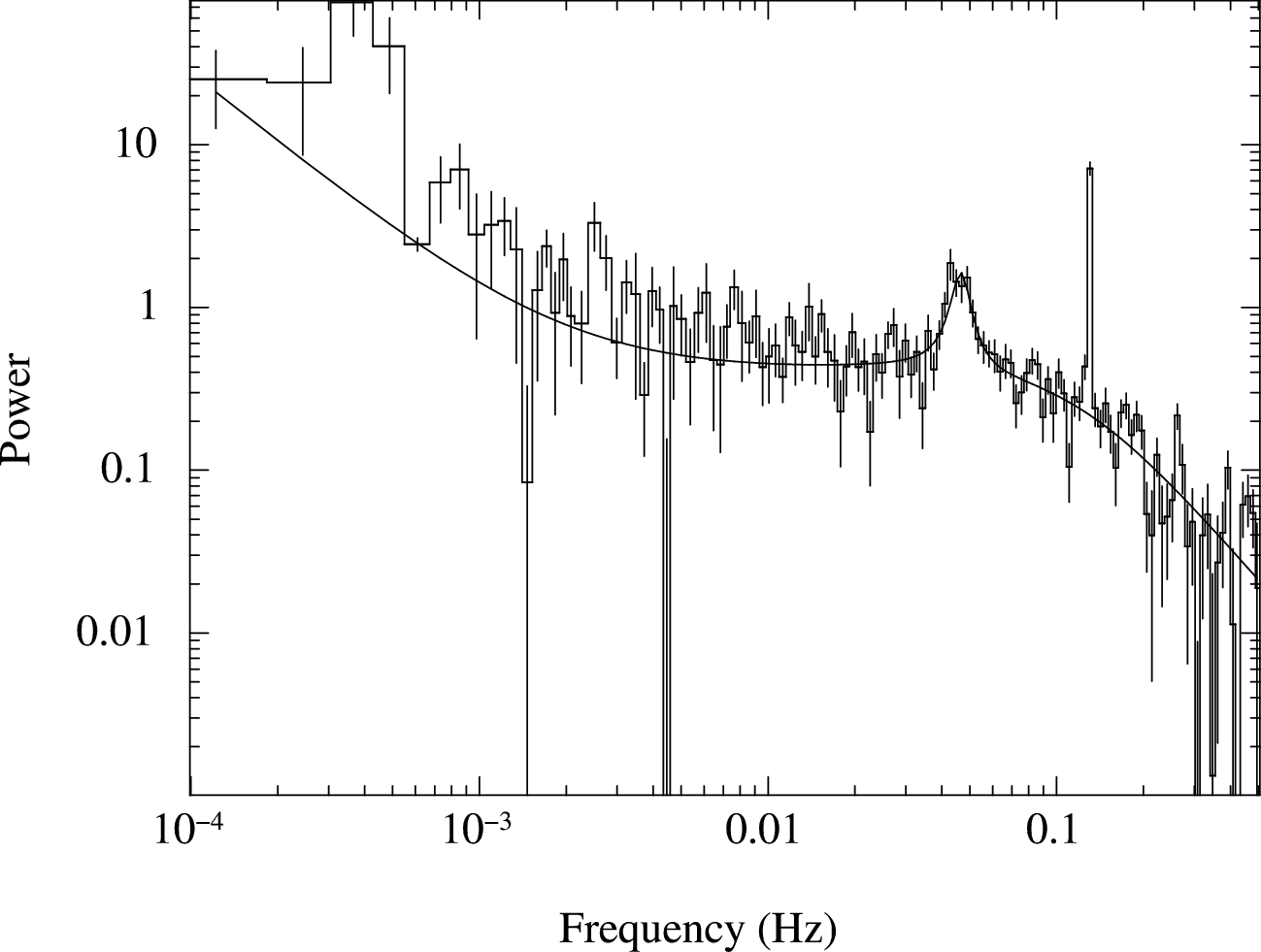}
    \caption{The PDS of 4U 1626-67 from {\it AstroSat}/LAXPC observation of Data 6 fitted with a Powerlaw and two Lorentzians.}
    \label{pds_fit}
\end{figure}

\begin{figure*}
   \includegraphics[width=0.45\textwidth, angle=0]{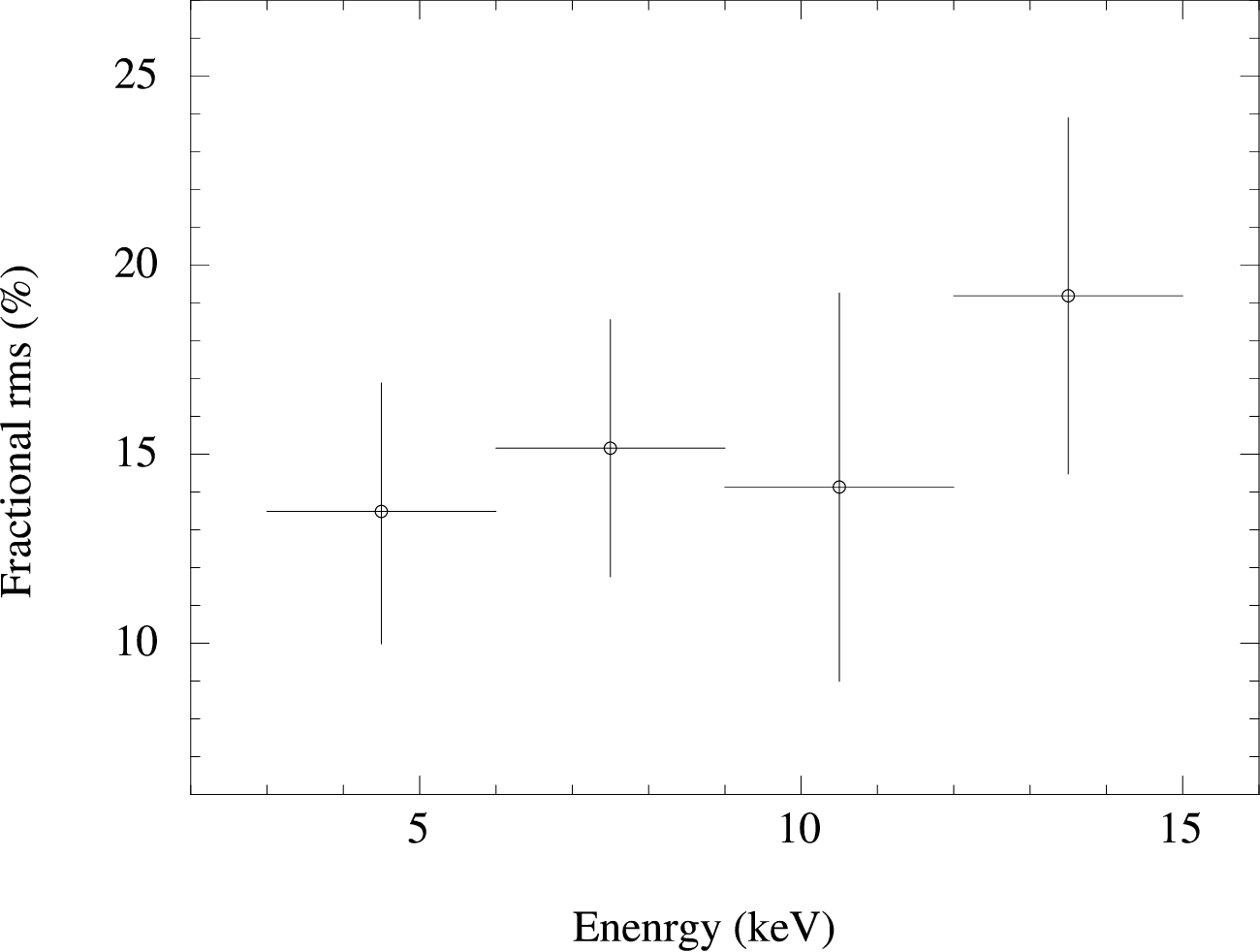}
   \hspace{1cm}
   \includegraphics[width=0.45\textwidth, angle=0]{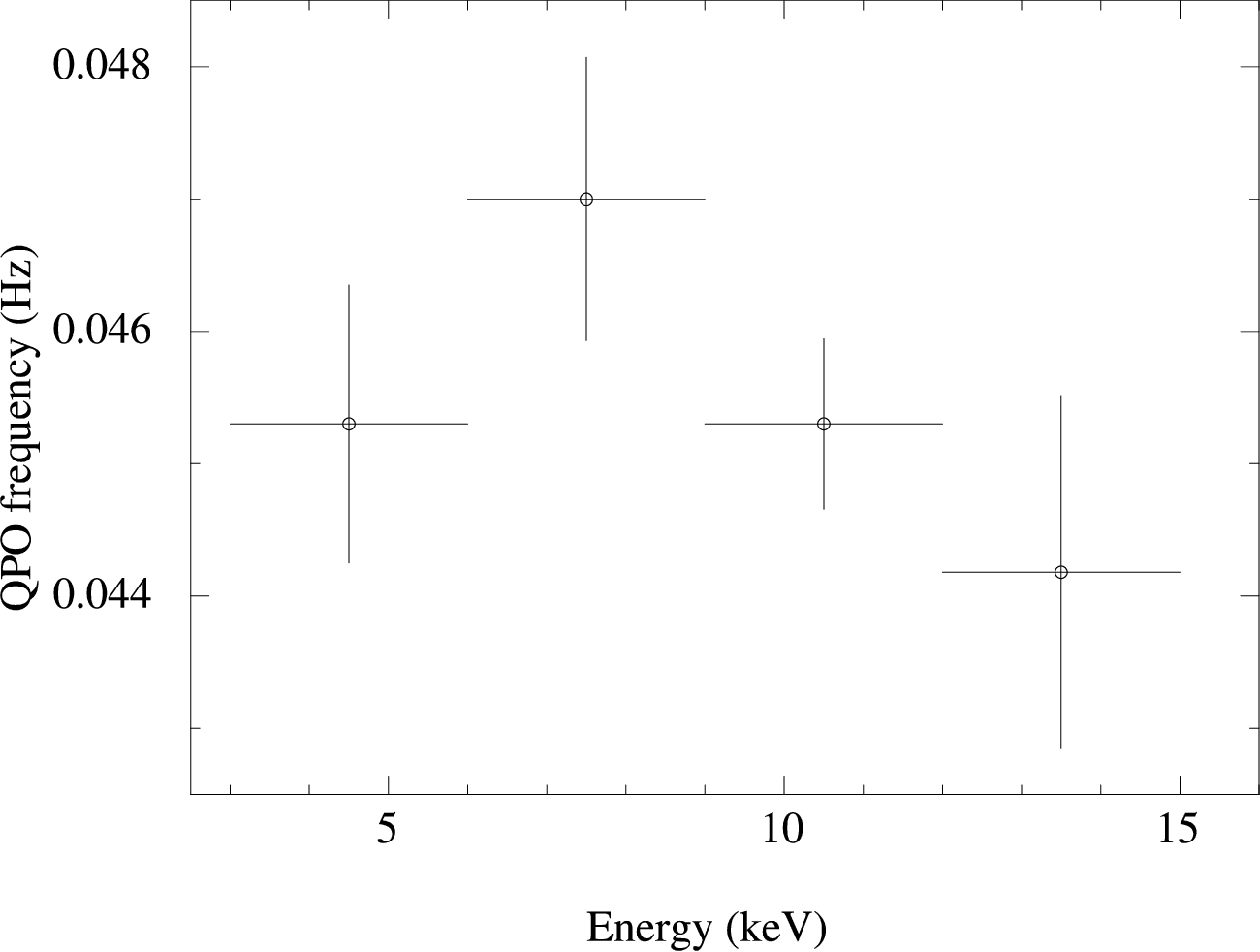}
   \caption{The energy-dependent variation of QPO fractional RMS amplitude (left) and QPO centre frequency (right) of 4U 1626-67 from {\it Astrosat}/LAXPC observations.} 
   \label{ene_rms_freq}
\end{figure*}

The continuum of the power spectrum of Data 6 was fitted with a model consisting of a power law component with a Photon Index (PI) of 1.03 and a Lorentzian with a Lorentzian Centre (LC) of 0.2 Hz. The QPO feature was fitted by adding another Lorentzian with the value of LC to be around 0.045 Hz. The fitted PDS is shown in Figure \ref{pds_fit}.
The quality factor of the QPO is about 2.96, and the QPO significance is 3.7 $\sigma$. The RMS fractional variability calculated from the background-subtracted data is $11.94\pm3.2 \%$. The centre frequency of the QPO is found to be $46.5\pm1.0$ mHz, and the QPO feature has a width of about $7.76\pm2.3$ mHz. 
The left panel of Figure \ref{ene_rms_freq} presents the energy dependence of the rms value of the QPO. While significant uncertainties in the results limit definitive conclusions, the observed positive correlation between rms value and energy is consistent with previous studies of the source \citep{manikantan2024}.
The change in QPO centre frequency with energy is studied and the plot is shown on the right panel of Figure \ref{ene_rms_freq}. 
The centre frequency is observed to increase up to approximately 8 keV, beyond which it exhibits a decrease.
We could not check the time evolution of the QPO frequency as the available data is insufficient for such an analysis. 

\subsection{Spectral analysis} \label{subsec:spec}

We used $XSPEC$ version 12.12.1 for the spectral fitting. We analysed the 3.0–20.0 keV spectra from four of the six LAXPC20 observations. Data 1 and 2 have very little exposure time and thus poor statistics; leading to the exclusion of the spectra of those observations from further analysis. 

In analysing the spectra of Data 3 to Data 5, we used the spectral model with a power law and blackbody. 
The absorption component was introduced by the model {\sc tbabs}\footnote{https://heasarc.gsfc.nasa.gov/xanadu/xspec/manual/node268.html}, which describes the X-ray absorption by the Inter-Stellar Medium (ISM). The absorption column density (n$_{H}$) value is fixed at $9.52\times10^{20}$ cm$^{-2}$, as given by \cite{HI4PI}. We did not detect any difference in the parameters of the blackbody component with or without freezing the Hydrogen column density. 
Adding Gaussian at the broad iron emission line detected, improved the fit and reduced the value of $\Delta\chi^{2}$. 
The iron line energies obtained in our results are higher than those reported by \cite{Tobrej2024MNRAS} and \cite{Sharma2023MNRAS}. This discrepancy could be due to the lower spectral resolution (FWHM ~1.2 keV at 6 keV) of LAXPC \citep{Antia2017ApJS}.

The final used spectral model is $tbabs*(powerlaw + bbody + gaussian)$. The spectral uncertainties of the parameter values are computed using the \textsl{error} command, to 90\% confidence level.
The detected value of the photon index was around 0.96. The blackbody temperature varied between 0.62 keV and 0.68 keV.
The observed flux value is decreasing over the years from $6.18\times10^{-10}$ erg cm$^{-2}$s$^{-1}$ in 2016 August to $5.54\times10^{-10}$ erg cm$^{-2}$s$^{-1}$ in 2017 July to $4.81\times10^{-10}$ erg cm$^{-2}$s$^{-1}$ in 2018 May. For all these spin-up state observations, the reduced chi-square values are lower than 1. However, the results remain consistent with previous studies \citep{Beri2014MNRAS,Beri2018MNRAS, Sharma2023MNRAS,Tobrej2024MNRAS}.

\begin{figure}
   \includegraphics[width=0.315\textwidth, angle=-90]{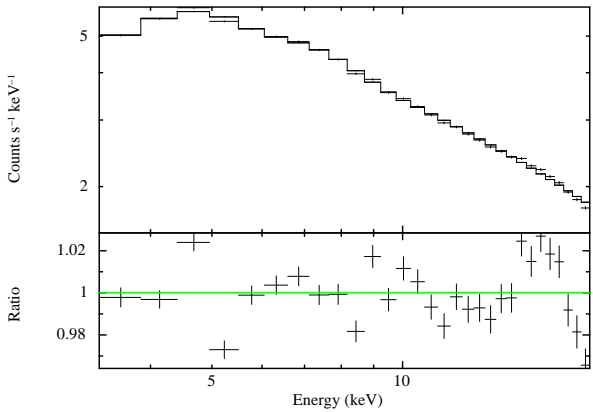}
    \caption{The best-fit spectrum and ratio of Data 5 to the model $tbabs*(powerlaw + bbody + gaussian)$ from {\it Astrosat}/LAXPC observation of 4U 1626-67.} 
   \label{spec_2018may}
\end{figure}

A spectral model with powerlaw and blackbody is used to fit the Data 6 spectrum. We also used the $gain$ command in $XSPEC$ to account for the uncertainty in the gain of the LAXPC20 detector. Even though no significant iron emission line was detected, a broad feature around $\sim$ 7 keV is seen in the spectra. But the addition of Gaussian did not improve $\Delta \chi^{2}$, thus we have used the model $tbabs*(powerlaw + bbody)$ for the spectral fit.  
The value of the photon index lowered to 0.85 and the blackbody temperature changed to 2.24 keV. The flux value also decreased to $1.75\times10^{-10}$ erg cm$^{-2}/$s$^{-1}$.
The best-fitting spectrum of the 2018 May observation is shown in Figure \ref{spec_2018may} and the best-fitting parameters of all the spectral observations are given in Table \ref{tab:spec}.

\begin{table*}
\begin{center}
\caption{ Best-fitting parameters of the spectra of 4U 1626-67 from Data 3 to Data 6.} \label{tab:spec}
    \begin{tabular}{lccccccr} 
\hline \\[2pt]
  Parameter & Data 3 & Data 4 & Data 5 & Data 6 \\[5pt]
\hline \\[2pt]

PhoIndex ($\Gamma\ $) & $0.99_{-0.05}^{+0.04}$ & $0.95\pm0.05$ & $0.96_{-0.05}^{+0.04}$ & $0.85_{-0.25}^{+0.19}$ \\[10pt]

Norm$_{PL}$ & $2.1\pm0.2\times10^{-2}$ & $1.7\pm0.2\times10^{-2}$ & $1.5_{-0.2}^{+0.1}\times10^{-2}$ & $2.83_{-1.5}^{+1.9}\times10^{-3}$ \\[10pt]

kT$_{Bbody}$ (keV) & $0.628_{-0.24}^{+0.22}$ & $0.68_{-0.2}^{+0.17}$ & $0.66_{-0.15}^{+0.16}$ & $2.24_{-0.08}^{+0.12}$ \\[10pt]

Norm$_{Bbody}$ & $9.78\pm4.0\times10^{-4}$ & $8.16_{-2.0}^{+9.0}\times10^{-4}$ & $6.22_{-2.0}^{+5.2}\times10^{-4}$  & $8.81_{-1.7}^{+1.9}\times10^{-4}$  \\[10pt]

LineE (keV)& $7.26_{-1.39}^{+0.54}$ & $7.2_{-1.2}^{+0.5}$ & $7.37_{-0.65}^{+0.36}$ & - \\[10pt]

Sigma (keV)& $1.0_{-0.71}^{+1.03}$ & $1.23_{-0.57}^{+0.87}$ & $0.92_{-0.59}^{+0.76}$ & - \\[10pt]

Eqw (keV) & $0.29_{-0.21}^{+0.26}$ & $0.44_{-0.43}^{+0.45}$ & $0.32_{-0.31}^{+0.33}$ & - \\ [10pt]

Norm$_{Gaus}$ & $8.86_{-4.0}^{+1.4}\times10^{-4}$ & $11.6_{-5.6}^{15.7}\times10^{-4}$ & $7.47_{-3.4}^{+7.7}\times10^{-4}$ & - \\[10pt]

Flux (erg cm$^{-2}$sec$^{-1}$) & $6.18_{-0}^{+0}\times10^{-10}$ & $5.54_{-0}^{+0}\times10^{-10}$ & $4.81_{-0}^{+0}\times10^{-10}$ & $1.75_{-0}^{+0}\times10^{-10}$ \\[10pt]

$\chi^{2}/$ dof & 14.28/21 & 15.08/23 & 14.68/23 & 21.62/24 \\[10pt]

\hline
\end{tabular}
\end{center}
\end{table*}

\section{Discussion}
\label{sect:discussion}

This work reports results of the timing and spectral analysis of the persistent pulsar 4U 1626-67 during its spin-up and spin-down states from January 2016 to May 2023, using {\it AstroSat}/LAXPC observations. Since its discovery, this source has experienced multiple torque reversals, including a transition from spin-up to spin-down in 1990, spin-down to spin-up in 2008, and a return to spin-down in 2023. Such torque reversals are observed in a few other sources like GX 1+4, Cen X-3, and Her X-1 \citep{Bildsten1997ApJS}. Among these, GX 1+4 and 4U 1626-67 have maintained decades-long torque states with steady spin-up and steady spin-down \citep{Perna2006ApJ}.  

A distinctive feature of the LMXB pulsar 4U 1626-67 is the presence of flares in its light curve, which are generally observed during periods of spin-up \citep{Joss1978ApJ, Beri2014MNRAS, Beri2018MNRAS}.
We detected strong flares, with intensity reaching two to three times that of the quiescent level, in the light curve when the source was in its second spin-up era. Flares with similar characteristics have been previously observed in this state in X-ray data \citep{Beri2014MNRAS, Beri2018MNRAS, Schulz2019arXiv} and in optical data \citep{Raman2016MNRAS}. 
The flares observed in our results exhibit a symmetric rise and decay, followed by a dip in persistent emission. This behaviour is similar to the Type II bursts in GRO J1744-28, which showed the presence of a dip and a recovery period following each outburst \citep{Giles1996ApJ}. The Lightman-Eardley (LE) instability \citep{Lightman1974ApJ} is suggested to be the dominant mechanism behind the Type II bursts in GRO J1744-28 \citep{Li1997A&A, Cannizzo1996ApJ}.
The LE instability arises when the radiation pressure in the inner accretion disc becomes comparable to the gas pressure, and the material that is evacuated onto the pulsar during an accretion event is replenished by material flowing in from further out. This causes the dip and recovery in the light curve following an outburst \citep{Cannizzo1996ApJ}. The presence of a dip following the two consecutive flares in Data 3 of our observation points to the possibility that an accretion instability—potentially the LE instability—contributes to the flaring behaviour of 4U 1626-67.
Previous XMM-Newton \citep{Beri2018MNRAS} and Chandra \citep{Schulz2019arXiv} observations of the source during its second spin-up phase also report dips following strong flares. Notably, \cite{Beri2018MNRAS} also suggested that the LE instability could be responsible for these events.
In our observation of Data 6, no flaring activity was detected, consistent with previous reports of minimal flaring during spin-down periods. This absence of flares may be due to localized instabilities within the accretion disc being absent or less pronounced due to the lower accretion rates during spin-down.

Consistent pulsations at $\sim$ 7.67 s were detected in all the {\it Astrosat}/LAXPC observations carried out in this work, during both torque states. Among these, Data 6 was characterized by the presence of a $46.5\pm1.0$ mHz QPO. We could detect the QPO feature up to $\sim$ 15 keV. QPOs around 46 mHz and 48 mHz have been detected in the X-ray emission from 4U 1626-67 \citep{Kommers1998ApJ, Chakrabarty2001ApJ, manikantan2024} in the spin-down eras and weak $\sim$ 40 mHz QPOs have been detected during the initial spin-up phase \citep{Shinoda1990PASJ}. The presence of QPOs typically during the spin-down state indicates changes in the source's accretion flow geometry with the torque reversal. It can be explained by assuming that QPOs are a manifestation of clumps in the inner accretion disc \citep{Beri2014MNRAS}.

QPOs are usually detected in 10-25 keV energy range in accretion-powered pulsars \citep{Raichur2008ApJ, James2010MNRAS}, with a notable difference in the case of GX 304-1 in which QPO feature is seen up to 40 keV \citep{Devasia2011MNRAS_GX304-1}. 
The energy dependence of the QPO centre frequency and fractional RMS of the QPO are examined. 
The centre frequency increases, peaking within the 3–9 keV energy range before declining. While the fractional RMS exhibits a positive trend, we cannot make a definitive conclusion due to the large associated uncertainties.
The correlation of QPO RMS with energy in 4U 1626-67 was previously observed by \citep{manikantan2024}. 
The observed positive correlation of the fractional RMS amplitude with energy could offer valuable insights into the physical mechanisms driving the QPOs. Studies by \cite{Gilfanov2003A&A} and \cite{Mukherjee2012ApJ} suggest that these oscillations may be due to the temperature fluctuations in a blackbody-like component.

The Keplerian frequency of the QPO is determined to be, $\nu_{k}$ = $\nu_{spin}$ + $\nu_{QPO}$ $\approx$ 0.17 mHz using the beat frequency model. BFM suggests that QPOs arise at the beat frequency between the neutron star’s spin frequency and the Keplerian frequency at the Alfvén radius of the disc-magnetosphere \citep{Alpar1985Nature}. This provides a consistent way to measure the magnetic dipole moment. The magnetic field strength of the pulsar can be approximated by the relation $B_{12} \sim$ 5.56 $\sqrt{L_{37}}$ \citep{Shinoda1990PASJ}. 
This measurement yields an estimate for the magnetic field strength of  $\sim$ 2.55$\times$10$^{12}$ Gauss. Previous studies have detected a Cyclotron Resonance Scattering Feature (CRSF) in the spectrum of this pulsar \citep{Orlandini1998ApJ}, indicating a surface magnetic field strength of $\sim$ 3$\times$10$^{12}$ Gauss. 
The magnetic field strength obtained from our study is within a reasonable range of this.

Our spin-up observations, which include detected flares, reveal an energy-dependent broad feature around $\sim$ 2 mHz in the power density spectra. Previous studies have reported QPO$-$like features around $\sim$ 3 mHz in the PDS generated from X-ray \citep{Joss1978ApJ, Beri2018MNRAS} and optical \citep{Raman2016MNRAS} data during previous spin-up states of this source. These studies suggest that such low-frequency features might be associated with the presence of flares during those observations.  
Although flares are evident in our observations, we cannot conclusively attribute the $\sim$ 2 mHz feature to them, as interpreting this feature with certainty is challenging due to its frequency being close to the harmonics of AstroSat's orbital frequency ($\sim$ 0.15 mHz) \citep{Antia2021JAPA}.

The broadening in the wings of the pulse frequency peak is observed during the spin-up state. This can be assumed to be due to the coupling between the periodic and aperiodic variabilities in the power spectra \citep{BURDERI1993Adv}.
It is also interesting to note that the broadening of the spin-frequency peak is not observed in the PDS where QPO is detected. The disappearance of the broadening during the presence of QPO has been previously reported in 4U 1901+03 by \cite{James2011MNRAS}.

A change in the shape of the power density spectra is noticed as the source underwent a torque reversal.
Torque reversals in disc-fed X-ray pulsars can plausibly be explained by the warping of the inner accretion disc, which may produce a retrograde disc flow close to the accreting neutron star \citep{van_Kerkwijk1998ApJ}.
Accretion from the inner disc to the neutron star surface likely occurs in different modes that are steady over a long time. Transitions between such stable accretion states that have different accretion flow geometry can result in different X-ray spectra and pulse characteristics \citep{van_Kerkwijk1998ApJ}. This might be the explanation behind the observed differences in pulse profile, spectra and power density spectra between the spin-up and spin-down states of 4U 1626-67.

The pulse profiles' energy and torque state dependency are seen in all the observations analysed in this work. In the spin-up state, the double-horned profile shifts to a single-peaked profile between lower ($<$ 12 keV) and higher ($>$ 18 keV) energies while the energy-averaged pulse profiles show a shoulder-like structure. The low energy profile structures are consistent with those seen in the spin-up eras of the pulsar in previous studies \citep{Beri2014MNRAS, Iwakiri2019ApJ}.
After the torque reversal, the pulse profiles in lower energies ($<$ 9 keV) show a flat top with multiple small peaks. Such profiles are obtained in the 2-12 keV energy range of the August 2003 XMM-Newton observation \citep{Koliopanos2016MNRAS} when the source was in its spin-down state. Higher energy ($>$ 15 keV) profiles have a broad sinusoidal structure with narrow spikes similar to the $>$ 18 keV profiles of the spin-up state.
The energy-averaged pulse profile in the spin-down state has a broad sinusoidal shape with many spikes; consistent with the previously reported observations \citep{Angelini1995ApJ, Krauss2007ApJ, Beri2014MNRAS}.

The difference in the shape of the pulse profile between spin-down and spin-up states can be explained using the change in the accretion flow geometry. The emission diagram of the accretion column depends on the mass accretion rate, transitioning between pencil-beam and fan-beam emission \citep{Basko1975A&A}. During the spin-down state, when the mass accretion rate is low, the inflowing material is irregular, forming a less stable, narrow accretion column {\citep{Iwakiri2019ApJ}}. This results in emission from the polar hot spots concentrated along the magnetic field axis and directed away from the accretion disc, leading to a pencil-beam emission pattern. Such a configuration is typically associated with single-peaked pulse profiles, though gravitational effects and various obscuration mechanisms can introduce more complex shapes \citep{Meszaros1992hermbook}.

When the mass accretion rate is high (spin-up state), the accretion column becomes filled with high-density plasma that slowly sinks under the neutron star’s gravitational field. This increases opacity along the magnetic field axis, causing most X-ray photons to escape through the optically thin sides of the accretion column, with the emission directed parallel to the accretion disc, forming a fan-beam pattern \citep{Koliopanos2016MNRAS}. This accounts for the narrow dips observed between the double peaks in the low-energy pulse profiles of the spin-up era \citep{White1983ApJ, Rea2004A&A}. Energy-dependent dips in pulse profiles have been observed in numerous X-ray pulsars, with these features being particularly prominent in the lower energy bands. \citep{Devasia2011MNRAS_1A1118-61, Maitra2012MNRAS}.
At higher energies, the absorption effect is reduced, allowing emission from the polar hot spots to be significant. The similarity of the higher energy profiles in the two torque states, also observed by \cite{Beri2014MNRAS}, suggests the presence of a high-energy emission component with a broad pulse profile.

A correlation between the spectral parameters and the torque state of 4U 1626-67 was revealed by fitting the spectrum with the widely accepted spectral model. During the observations of the spin-up era, the time-averaged spectrum was well fitted by a blackbody temperature kT of $\sim$ 0.6 keV and a power law photon index of $\sim$ 0.96. Similar spectral parameters are obtained for the source in previous spin-up states \citep{Pravdo1979ApJ, Kii1986PASJ, Koliopanos2016MNRAS, Beri2018MNRAS}. 
In the spin-down state, the photon index lowered to $\sim$ 0.85 suggesting a hardening of the spectra, which is also observed in the earlier spin-down states. When compared to the previous spin-down state observations, we obtained a higher value of $kT_{bb}$ although a lower value was expected. A previous study of the source using the {\it NuSTAR} observation, coinciding with our observational period, did not detect a blackbody component in their spectra \citep{Sharma2023MNRAS}. Whereas, a recent study of both {\it NuSTAR} and {\it NICER} on the same observational period have revealed a blackbody temperature of $\sim$ 0.25 keV \citep{Tobrej2024MNRAS}.

The flux value obtained from our observations decreased as 4U 1626-67 went from the spin-up era to the spin-down era. A similar trend of lowering flux value is noted in the current and previous spin-down states of the source  \citep{Jain2010MNRAS, Sharma2023MNRAS}.
The luminosity of the source obtained in the spin-up state varies between 4.66$\times$10$^{36}$ erg s$^{-1}$ and 5.99$\times$10$^{36}$ erg s$^{-1}$. 
After the torque reversal, the source luminosity changed to 2.1$\times$10$^{36}$ erg s$^{-1}$, indicating that the average luminosity of the source is in the order of 10$^{36}$ erg s$^{-1}$. If the luminosity of the accreting pulsars is in the range L $\leq$ 10$^{35-37}$ erg s$^{-1}$, then the emission pattern may be characterised by the combination of a pencil beam and a fan-beam \citep{Kraus2003apj}. 
This along with the difference in the shape of the pulse profiles and the presence and absence of QPOs can be explained by the change in the accretion flow geometry of 4U 1626-67 between spin-up and spin-down states.

\normalem
\begin{acknowledgements}
This research is based on the results obtained from the {\it AstroSat} mission of the Indian Space Research Organization (ISRO), archived at the Indian Space Science Data Centre (ISSDC). 
This work uses data from the LAXPC instrument. We thank the LAXPC Payload Operation Center (POC) at TIFR, Mumbai, for providing the data and the necessary software tools. 
We thank Prof. Indulekha Kavila for her insightful comments and constructive suggestions, particularly in the discussion section of this paper. Author Juris thanks Prof. H.M.Antia for his valuable clarifications through email correspondence.
Authors Juris and Marykutty acknowledge the financial support from ISRO (Sanction Order:No.$DS\_2B-13013(2)/3/2021-Sec.2$).
We thank the anonymous referee for their insightful remarks and recommendations, which helped to improve this work. 
\end{acknowledgements} 
  
\bibliographystyle{raa}
\bibliography{bibtex}

\end{document}